# Line geometry and electromagnetism IV:
# electromagnetic fields as infinitesimal Lorentz transformations

D. H. Delphenich ([†])
Spring Valley, OH 45370, USA

**Abstract.** It is first shown that the scalar product on any orthogonal space $(V, g)$ allows one to define linear isomorphisms of the vector spaces of bivectors and 2-forms on $V$ with the underlying vector spaces of the Lie algebra $\mathfrak{so}(p, q)$ and its dual, respectively. When those isomorphisms are applied to the electromagnetic excitation bivector and field strength 2-form, resp., one can associate various algebraic constructions that pertain to them as bivector fields and 2-forms with corresponding constructions in terms of $\mathfrak{so}(1, 3)$ and its dual. The subsequent association with corresponding things in line geometry will then become straightforward. In particular, the fields can be represented by "motors," such as "screws and wrenches," while the Cartan-Killing form on $\mathfrak{so}(1, 3)$ is isometric to the scalar product on $\Lambda_2$ that gives the Klein quadric. When $\Lambda_2$ (and therefore $\Lambda^2$) is given an almost-complex structure (and therefore, a complex structure), one can also represent most of the constructions on $\Lambda_2$ and its dual in terms of $\mathfrak{so}(3; \mathbb{C})$ and its dual.

## Table of contents

———

**1. Introduction.** – This article is a continuation of a series of papers [1] that have each examined the extent to which one can think of line geometry as the geometry of electromagnetism in the same way that metric geometry is the geometry of gravitation. Hence, a certain familiarity with the elements of the subject will be assumed, although some attempt has been made to make the discussion here relatively self-contained.

If one goes back to the basic definitions of the electric field strength **E** and magnetic field strength **H** then one must recall that they are defined in terms of dynamical concepts. That is, **E** represents the force **F** on a unit charge $q$, or more precisely:

$$\mathbf{E} = \lim_{q \to 0} \frac{1}{q} \mathbf{F}(q) \,. \tag{1.1}$$

---

([†]) Website: neo-classical-physics.info. E-mail: feedback@ neo-classical-physics.info.

As for the magnetic field strength, one actually has a choice: Typically, one thinks of **B** as something that exerts a *force* on a unit *current* **I**, as the Lorentz force law ([1]):

$$\mathbf{F} = \mathbf{I} \times \mathbf{B} \tag{1.2}$$

would suggest, or rather, the corresponding limit as the current goes to zero.

However, since a magnetic field **B** also exerts a *torque* on a magnetic dipole $\boldsymbol{\mu}$:

$$\boldsymbol{\tau} = \boldsymbol{\mu} \times \mathbf{B}, \tag{1.3}$$

and a magnetic dipole is (as far as any experimenter has managed to show) the most elementary source of a magnetic field, one can also think of (the magnitude of) **B** as the limiting (magnitude of the) torque on a magnetic dipole $\boldsymbol{\mu}$ as its strength $\mu$ goes to zero:

$$B = \lim_{\mu \to 0} \frac{1}{\mu} \tau(\mu). \tag{1.4}$$

Once again the direction of **B** will be perpendicular to that of $\boldsymbol{\tau}$.

The reason that one might wish to go the latter route in one's basic definitions is that bivectors and 2-forms can be used for representing kinematical and dynamical concepts, in such a way that a bivector can represent an infinitesimal rigid motion in three-dimensional Euclidian space (which is a clearly non-relativistic notion) or an infinitesimal Lorentz transformation. Similarly, a 2-form can represent an element of the dual vector space of either of the aforementioned Lie algebras. When one looks at the time+space decomposition of the electromagnetic bivector field $\mathfrak{H}$ or the 2-form *F*, one will see that it is more natural to associate electric fields with forces and magnetic fields with torques.

One also has a choice in terms of how one regards the vector cross product on $\mathbb{R}^3$:

1. Due to its antisymmetry, one can think of it as a precursor to the more general exterior product that takes advantage of certain "accidental" isomorphisms of vector spaces that are true only for three dimensions.

2. One can regard it as defining a Lie algebra on $\mathbb{R}^3$, namely, the Lie algebra $\mathfrak{so}(3)$.

In this article, we will not choose one or the other option, but take advantage of the fact that the linear isomorphism that associates bivectors with elements of orthogonal Lie algebras is more general than it appears for a three-dimensional Euclidian space. Hence, we will be considering both aspects of the cross product.

The second section of this article will then exhibit that isomorphism in its general form and then specialize it to the cases that are relevant to the electromagnetic fields on four-dimensional Minkowski space. The third section will then examine that very association of electromagnetic fields with corresponding elements of $\mathfrak{so}(1, 3)$ and its dual vector space. In the fourth section, that association will then be combined with the association of bivectors and 2-forms with objects in line geometry that has been the main focus of this series of papers, and in the final section, the main results of this article will be summarized.

---

([1])   Of course, the direction of the force will be perpendicular to the direction of **B** as a result of this.

**2. Representing $\mathfrak{so}(1, 3)$ by bivectors.** – The basic fact that we shall elaborate upon in this section is that when you lower an index of the components $\omega^{\mu\nu}$ of a bivector $\boldsymbol{\omega} \in \Lambda^2 V$ for some frame on an $n$-dimensional orthogonal space $(V, g)$ (or raise the index of the components $\omega_{\mu\nu}$ of a 2-form, resp.) by using the metric $g_{\mu\nu}$ on $V$ then, due to the antisymmetry of $\omega$ in the index-pair $\mu\nu$, the resulting matrix:

$$\omega_\mu^{\cdot\nu} = g_{\mu\kappa}\,\omega^{\kappa\nu} \qquad (\omega^\mu_{\cdot\nu} = g^{\mu\kappa}\,\omega_{\kappa\nu}, \text{ resp.}) \qquad (2.1)$$

will belong to $\mathfrak{so}(p, q)$ [its dual $\mathfrak{so}(p, q)^*$, resp.]. Here, we are assuming that the signature type of the metric $g$ is $(p, q)$; hence, in an orthonormal frame on $V$, one will have:

$$g_{\mu\nu} = \eta_{\mu\nu} = \text{diag}\,[+1, \ldots, +1, -1, \ldots, -1], \qquad (2.2)$$

with $p$ plus signs and $q$ minus signs.

After that, since the association is a linear isomorphism of vector spaces, one can transfer all of the algebraic machinery on $\mathfrak{so}(p, q)$ to $\Lambda^2 V$ in a manner that makes many of the common constructions in terms of bivectors and 2-forms take on a corresponding significance in terms of the Lie algebra of infinitesimal orthogonal transformations for the chosen scalar product.

Since this is essentially a corollary to a more general result that involves $\mathfrak{gl}(n)$, we shall start at that more "pre-metric" level of multilinear algebra [**2**].

*a. Representing* Hom $V$ *and its dual by second-rank tensors.* – The basic fact that we shall start with is that if $V$ is an $n$-dimensional vector space (whether real or complex) and $V^*$ is its dual then the elements of the tensor product $V^* \otimes V$ can be associated with linear transformations of $V$ to itself. If one calls that vector space of linear transformations Hom $V$ then the way that one can define the linear isomorphism $V^* \otimes V \cong$ Hom $V$ is to first note that if $\alpha \otimes \mathbf{v}$ is a decomposable element of $V^* \otimes V$ then the map $\alpha \otimes \mathbf{v} : V \to V$ that takes any vector $\mathbf{x} \in V$ to the vector $\alpha(\mathbf{x})\,\mathbf{v}$ is linear since $\alpha$ is linear. However, not all elements of $V^* \otimes V$ are decomposable, but more generally they are finite linear combinations of decomposable elements. Hence, one "extends the map by linearity" to all finite linear combinations of decomposable elements in order to get the full isomorphism. That is, if $T = \sum_i \alpha^i \otimes \mathbf{v}_i$ is a finite linear combination of decomposable elements then the action of $T$ on a vector $\mathbf{x}$ will be:

$$T(\mathbf{x}) = \sum_i \alpha^i(\mathbf{x}) \, \mathbf{v}_i \, . \tag{2.3}$$

The isomorphism is, perhaps, easier to understand when one looks at components of the tensors in $V^* \otimes V$ and the matrices of the corresponding linear transformations of Hom $V$. Suppose $\{\mathbf{e}_i, i = 1, \ldots, n\}$ is a frame (i.e., basis) for $V$, and $\{\theta^i, i = 1, \ldots, n\}$ is its reciprocal basis for $V^*$; hence:

$$\theta^i(\mathbf{e}_j) = \delta^i_j \, . \tag{2.4}$$

One can then use $\{\theta^i \otimes \mathbf{e}_j, i, j = 1, \ldots, n\}$ as a frame for $V^* \otimes V$, and any element $T$ of that vector space can be expressed in the form:

$$T = T_i{}^{\cdot j} \, \theta^i \otimes \mathbf{e}_j \, . \tag{2.5}$$

However, every linear transformation $T : V \to V$ can also be associated with a matrix $T_i{}^{\cdot j}$ by setting:

$$T(\mathbf{e}_i) = T_i{}^{\cdot j} \, \mathbf{e}_j \, . \tag{2.6}$$

From the standpoint of component matrices, the isomorphism that we are discussing becomes the direct association of $T_i{}^{\cdot j}$, the tensor component matrix, with $T_i{}^{\cdot j}$, the matrix of the linear map, which is essentially an identity at the level of components.

The dual isomorphism $V \otimes V^* \cong (\text{Hom } V)^* \cong \text{Hom } V^*$ is obtained by starting with the action of a decomposable element $\mathbf{v} \otimes \alpha \in V \otimes V^*$ on a linear functional $\beta \in V^*$ to produce $\beta(\mathbf{v}) \, \alpha$ and extending by linearity. As far as components are concerned, this time, a tensor $T \in V \otimes V^*$ will take the form:

$$T = T_{\cdot j}^{i} \, \mathbf{e}_i \otimes \theta^j, \tag{2.7}$$

and the matrix of a linear map $T : V \to V$ will be obtained from:

$$T(\theta^i) = T_{\cdot j}^{i} \, \theta^j, \tag{2.8}$$

so the association of the tensor component matrix with the matrix of the linear transformation will once more be the one that takes $T_{\cdot j}^{i}$ to itself.

*b. The bilinear pairing of* Hom $V^*$ *with* Hom $V$. – The natural bilinear pairing:

$$\text{Hom } V^* \times \text{Hom } V \to \mathbb{K}, \qquad (T^*, S) \mapsto T^*(S) \tag{2.9}$$

(in which the scalar field $\mathbb{K}$ is either $\mathbb{R}$ or $\mathbb{C}$, for us) corresponds to the bilinear pairing of matrices:

$$(T^{i}_{\cdot j}, S^{\cdot l}_{k}) \mapsto T^{i}_{\cdot j} S^{\cdot j}_{i} = \text{Tr } [T][S], \qquad (2.10)$$

in which we have abbreviated the matrices to $[T]$ and $[S]$.

*c. The algebras that can be defined on second-rank tensors.* – The vector space Hom $V$ has another natural structure beyond its vector space structure, namely, a bilinear binary operation Hom $V \times$ Hom $V \to$ Hom $V$, $(S, T) \mapsto ST$ that is defined by the composition of linear maps to give linear maps. That bilinear binary operation then gives Hom $V$ the structure of an *algebra*. In order to distinguish the vector space Hom $V$ from the algebra that it defines, we shall denote the latter by $\mathfrak{hom}(n)$.

One can use the existence of the linear isomorphism $V^* \otimes V \cong$ Hom $V$ to define a corresponding algebra on $V^* \otimes V$. For our purposes, it is easiest to describe in terms of matrices, since it basically amounts to matrix multiplication for the component matrices of tensors in $V^* \otimes V$. One is then dealing with the bilinear pairing:

$$(S^{\cdot j}_{i}, T^{\cdot l}_{k}) \mapsto S^{\cdot k}_{i} T^{\cdot j}_{k} \; . \qquad (2.11)$$

One can then polarize this algebra product by commutation:

$$ST = \tfrac{1}{2} \{S, T\} + \tfrac{1}{2} [S, T], \qquad (2.12)$$

in which

$$\{S, T\} = ST + TS, \qquad [S, T] = ST - TS. \qquad (2.13)$$

Both of these brackets define algebra products by themselves, and we shall denote the algebra on Hom $V$ that the commutator bracket [.,.] defines by $\mathfrak{gl}(n)$, since it is the general linear Lie algebra. Its elements are infinitesimal generators of invertible linear transformations of $V$.

Typically, since the dual space (Hom $V)^*$ does not have a *natural* algebra product, one does not define one. Of course, if one has a linear isomorphism of $V$ with $V^*$ then one can define a corresponding linear isomorphism of Hom $V$ with its dual and give the latter vector space the induced algebra product, but that is not a natural construction, since it depends upon the choice of isomorphism of $V$ with $V^*$.

*d. Lowering and raising the indices.* – Suppose that one does have such a linear isomorphism $C : V \to V^*$, $\mathbf{v} \mapsto \mathbf{v}^*$. One can then define a bilinear functional on $V$:

$$C (\mathbf{v}, \mathbf{w}) = \mathbf{v}^*(\mathbf{w}). \qquad (2.14)$$

Dually, one has the inverse isomorphism $C^{-1} : V^* \to V$, $\alpha \mapsto \alpha^{*-1}$, which then defines a bilinear functional on $V^*$:

$$C^*(\alpha, \beta) = \beta(\alpha^{*-1}). \qquad (2.15)$$

In terms of components, if the matrix of $C$ relative to the aforementioned choice of frame and coframe is $C_{ij}$ then:

$$v_i \equiv (\mathbf{v}^*)_i = C_{ji} \, v^j, \qquad C(\mathbf{v}, \mathbf{w}) = C_{ij} \, v^i \, w^j. \qquad (2.16)$$

Dually, the matrix of the inverse isomorphism is $C^{ij}$, and:

$$\alpha^i \equiv (\alpha^{*-1})^i = C^{ji} \, \alpha_j, \qquad C^*(\alpha, \beta) = C^{ij} \, \alpha_i \, \beta_j. \qquad (2.17)$$

One can then combine the isomorphisms that were just defined with the ones above and get isomorphisms:

$$V \otimes V \cong V^* \otimes V \cong \mathrm{Hom}\, V, \quad V^* \otimes V^* \cong V \otimes V^* \cong \mathrm{Hom}\, V^*. \qquad (2.18)$$

From the standpoint of components, these isomorphisms amount to lowering and raising one index of the component matrix, respectively:

$$T^{ij} \mapsto T_i^{\cdot j} = C_{ki} \, T^{kj}, \quad T_{ij} \mapsto T_{\cdot j}^i = C^{ki} \, T_{kj}. \qquad (2.19)$$

(We have suppressed the second isomorphism and its dual, since they amount to identities when one looks at component matrices.)

One can also define the multiplication of elements in $V \otimes V$ in a manner that will correspond to the multiplication of elements in $\mathrm{Hom}\, V$; i.e., matrix multiplication. Hence, if:

$$S = S^{ij} \, \mathbf{e}_i \otimes \mathbf{e}_j, \qquad T = T^{ij} \, \mathbf{e}_i \otimes \mathbf{e}_j$$

then the components of $ST$ will be:

$$(ST)^{ij} = C^{lj} \, S_k^{\cdot i} \, T_l^{\cdot k} = C_{kl} \, S^{li} \, T^{jk}. \qquad (2.20)$$

One can put this into a form that does not involve the components by forming:

$$ST = (ST)^{ij} \, \mathbf{e}_i \otimes \mathbf{e}_j = C_{kl} \, S^{li} \, T^{jk} \, \mathbf{e}_i \otimes \mathbf{e}_j = C_{kl} \, (S^{li} \, \mathbf{e}_i) \otimes (T^{jk} \, \mathbf{e}_j),$$

which will become:

$$ST = C_{kl} \, (i_{\theta^l} S) \otimes (T_{\theta^k} i) \qquad (2.21)$$

when we define the left and right interior products by the coframe elements by:

$$i_{\theta^l} S = S^{ij} \, \theta^l(\mathbf{e}_i) \, \mathbf{e}_j = S^{lj} \, \mathbf{e}_j, \qquad T_{\theta^k} i = T^{ij} \, \mathbf{e}_i \, \theta^k(\mathbf{e}_j) = T^{jk} \, \mathbf{e}_j. \qquad (2.22)$$

Of course, the comment that was made at the end of the last subsection still applies: This construction of an induced multiplication on bivectors is not natural, since it depends upon the choice of isomorphism $C$. However, one sees that the product $ST$ in (2.21) will be invariant under any change of coframe $\overline{\theta}^i = A^i_j \theta^j$, since $C_{kl}$ will transform contragrediently to $\theta^k$:

$$\overline{C}_{ij} = \tilde{A}^k_i \, \tilde{A}^l_j C_{kl}. \tag{2.23}$$

Another isomorphism of vector spaces that is canonical and simple to explain in terms of components is the isomorphism of $V^* \otimes V$ with $V \otimes V^* = (V^* \otimes V)^*$. One starts by associating every decomposable element $\alpha \otimes \mathbf{v} \in V^* \otimes V$ with the corresponding element $\mathbf{v} \otimes \alpha \in V \otimes V^*$ and then extends to all finite linear combinations of decomposable elements by linearity. In particular, if $\theta^i \otimes \mathbf{e}_j$ gives a basis for $V^* \otimes V$ then $\mathbf{e}_j \otimes \theta^i$ will give a basis for $V \otimes V^*$, and the element $T = T_i^{\cdot j} \, \theta^i \otimes \mathbf{e}_j \in V^* \otimes V$ will go the element:

$$T^{\mathrm{T}} = T_i^{\cdot j} \mathbf{e}_j \otimes \theta^i = T^i_{\cdot j} \mathbf{e}_i \otimes \theta^j, \tag{2.24}$$

in $V \otimes V^*$. This shows that the linear isomorphism that we have defined essentially amounts to the transposition of a linear operator $L : U \to W$ to obtain a linear operator $L^{\mathrm{T}}$: $W^* \to U^*$ whose component matrix will then be the transpose of the component matrix of $L$.

*e. The specialization to $\mathfrak{so}(p, q)$.* – So far, nothing has been said about the symmetry of components. In particular, suppose that the isomorphism $C$ makes the corresponding bilinear form symmetric:

$$C\,(\mathbf{v}, \mathbf{w}) = C\,(\mathbf{w}, \mathbf{v}). \tag{2.25}$$

$C$ will then define a scalar product on $V$, while $C^{-1}$ will define a scalar product on $V^*$. In order to make things more familiar to the people who deal in the Lorentzian structure on space-time, we then replace the symbol $C$ with $g$, and introduce the notation $\eta_{ij}$ such that when the frame $\mathbf{e}_i$ is orthonormal, one will have:

$$g\,(\mathbf{e}_i, \mathbf{e}_j) = \eta_{ij} = \mathrm{diag}\,[+1, \ldots, +1, -1, \ldots, -1], \tag{2.26}$$

in which there are $p$ plus signs and $q$ negative signs. Thus, the group of orthogonal transformations of the orthogonal space $(V, g)$ will be $O\,(p, q)$, and the ones that preserve volume elements on $V$, as well, will be $SO(p, q)$. The corresponding Lie algebra of infinitesimal orthogonal transformations will be $\mathfrak{so}(p, q)$, in either case.

The condition for a transformation $T$ to be orthogonal – so $T \in O(p, q)$ – is that for every pair of vectors $\mathbf{v}, \mathbf{w}$ in $V$, one must have:

$$g\,(T(\mathbf{v}), T(\mathbf{w})) = g\,(\mathbf{v}, \mathbf{w}). \tag{2.27}$$

In terms of components, this will become:

$$T_i^k \, T_j^l \, g_{kl} = g_{ij}. \tag{2.28}$$

In order to get the corresponding condition for an infinitesimal transformation $t$ to be orthogonal – i.e., $t \in \mathfrak{so}(p, q)$ – one assumes that the matrix $T_j^{\ i}$ belongs to a differentiable curve $T_j^{\ i}(s)$ in $\mathfrak{gl}(n)$ through the identity matrix, so $T_j^{\ i}(0) = \delta_j^i$. Furthermore, we let:

$$t_j^{\ i} = \frac{d}{ds}\bigg|_{s=0} T_j^{\ i}(s). \qquad (2.29)$$

If we differentiate (2.28) at $s = 0$ then we will get:

$$g_{kl}\, t_i^{\ k}\, \delta_j^l + \, g_{kl}\, \delta_i^k\, t_j^{\ l} = 0,$$

since we are not varying $g_{ij}$, and this will simplify to:

$$g_{kj}\, t_i^{\ k} + g_{ki}\, t_j^{\ k} = 0, \qquad (2.30)$$

or

$$t_{ji} + t_{ij} = 0. \qquad (2.31)$$

The last condition says that $t_{ij}$ is an antisymmetric matrix, so $t_{ij}$ can also be used as the component matrix for a 2-form:

$$t = \tfrac{1}{2} t_{ij}\, \theta^i \wedge \theta^j. \qquad (2.32)$$

The dual of the argument that led from (2.27) to (2.32) that starts with the condition for an orthogonal transformation of $V^*$, which will give:

$$T_k^{\ i}\, T_l^{\ j}\, g^{kl} = g^{ij}, \qquad (2.33)$$

will conclude with the bivector:

$$\mathbf{t} = \tfrac{1}{2} t^{\ ij}\, \mathbf{e}_i \wedge \mathbf{e}_j. \qquad (2.34)$$

Putting all of this together, we get:

**Theorem:**

*For any orthogonal space $(V, g)$, there are linear isomorphisms:*

$$\Lambda_2 \cong \mathfrak{so}(p, q), \quad \Lambda^2 \cong \mathfrak{so}(p, q)^*.$$

**Proof:**

One needs only to make the $t$ matrices more specific. In the first case, one has $t_i^{\ \cdot j}$, and in the second, one has $t_{\cdot j}^{\ i}$. The first isomorphism lowers the index $i$ in $t^{\ ij}$, while the second one raises the $i$ in $t_{ij}$.

One can also think of the isomorphism $\Lambda^2 \cong \mathfrak{so}(p, q)^*$ as being essentially the transpose of the isomorphism $\Lambda_2 \cong \mathfrak{so}(p, q)$. That is, if $\iota : \Lambda_2 \to \mathfrak{so}(p, q)$ is the latter linear isomorphism then its transpose will be the linear isomorphism $\iota^T : \mathfrak{so}(p, q)^* \to (\Lambda_2)^* = \Lambda^2$; hence, the isomorphism $\Lambda^2 \cong \mathfrak{so}(p, q)^*$ will be just the inverse of that.

**Corollary:**

a) *One can define a Lie bracket on $\Lambda_2$ that makes the components of $[s, t]$ equal to:*

$$[s, t]^{ij} = g_{kl} (s^{li} t^{jk} - t^{ik} s^{lj}) = -g_{kl} (s^{li} t^{kj} - t^{ki} s^{lj}). \tag{2.35}$$

b) *One then has:*

$$[s, t] = -g_{kl}(i_{\theta^k} s \wedge i_{\theta^l} t). \tag{2.36}$$

**Proof:**

a) This is an anti-symmetrization of (2.20).

b) This is an anti-symmetrization of (2.21), when one takes into account the symmetry of $g_{kl}$ and the anti-symmetry of $t$, which makes:

$$t_{\theta^k} i = t^{jk} \mathbf{e}_j = -t^{kj} \mathbf{e}_j = -i_{\theta^k} t.$$

The basic effect of our theorem is that it allows one to think of a bivector as a representative of an infinitesimal orthogonal transformation and a 2-form as a representative of the dual to such a thing.

**Corollary:**

*The bilinear pairing of 2-forms and bivectors corresponds to the bilinear pairing of $\mathfrak{so}(p, q)^*$ with $\mathfrak{so}(p, q)$.*

**Proof:**

$$F_{ij} G^{ij} = F_i^{\cdot j} G_{\cdot j}^i. \tag{2.37}$$

Of course, since the indices of both $F$ and $G$ are antisymmetric, one will have:

$$F_i^{\cdot j} = -F_{\cdot j}^i, \qquad G_{\cdot j}^i = -G_i^{\cdot j}, \tag{2.38}$$

so one can also say that:

$$F_{ij} G^{ij} = F_{\cdot j}^i G_i^{\cdot j}. \tag{2.39}$$

The bilinear form:

$$<a, b> = \text{Tr ad } a \text{ ad } b \tag{2.40}$$

can be defined on any Lie algebra $\mathfrak{g}$, and it is referred to as the *Cartan-Killing form*. In this, ad $a$ is the linear transformation ad $a : \mathfrak{g} \to \mathfrak{g}$ that takes any $b \in \mathfrak{g}$ to:

$$(\text{ad } a)(b) = [a, b]. \tag{2.41}$$

By choosing a basis $\{\boldsymbol{\varepsilon}_a, \ a = 1, \dots, N\}$ for $\mathfrak{g}$, one can associate ad $a$ with a matrix $[\text{ad } a]^a_b$:

$$[\text{ad } a](\boldsymbol{\varepsilon}_b) = [\text{ad } a]^a_b \, \boldsymbol{\varepsilon}_a, \tag{2.42}$$

and, in fact:

$$[\text{ad } a]^a_b = c^a_{cb} \, a^c = - \, c^a_{bc} \, a^c, \tag{2.43}$$

in which the scalar constants $c^a_{bc}$ are the structure constants of $\mathfrak{g}$ for the chosen basis:

$$[\boldsymbol{\varepsilon}_a, \ \boldsymbol{\varepsilon}_b] = c^c_{ab} \, \boldsymbol{\varepsilon}_c. \tag{2.44}$$

If we abbreviate the notation $[\text{ad } a]^a_b$ to simply $a^a_b$ then the Cartan-Killing form will look like:

$$<a, b> = a^a_b \, b^b_a, \tag{2.45}$$

which is clearly the same form as (2.37).

Although the bilinear form $<a, b>$ is symmetric (from a basic property of the trace), it does not have to be non-degenerate; i.e., it does not have to define a scalar product on $\mathfrak{g}$. However, it is a basic theorem of Lie algebras [**3, 4**] that it will be non-degenerate iff $\mathfrak{g}$ is semi-simple; i.e., it has no non-trivial Abelian ideals.

For example, that is true for $\mathfrak{so}(3)$. In fact, if one regards $\mathfrak{so}(3)$ as $\mathbb{R}^3$ with the vector cross product then the Cartan-Killing form will amount to the Euclidian dot product on $\mathbb{R}^3$. It is also true for $\mathfrak{so}(1, 3)$, and we shall see that the Cartan-Killing form on $\mathfrak{so}(1, 3)$ is isometric to the scalar product on $\Lambda_2$ that gives one the Klein quadric.

**3. Application to electromagnetic fields.** – In the case of electromagnetic fields, one has the Minkowski field strength 2-form $F$ and the excitation bivector field $\mathfrak{H}$. Typically, one also assumes that there is a Lorentzian structure $g$ on the space-time manifold $M$ ([1]), which we assume to have the signature type $(1, 3)$; i.e., $\eta = \text{diag}[1, -1, -1, -1]$. Hence, the relevant orthogonal group will be $O(1, 3)$, and its Lie algebra will be $\mathfrak{so}(1, 3)$.

---

([1]) Naturally, that statement will no longer be true when one deals with the pre-metric form of electromagnetism.

*a. The action of the Lorentz group on bivectors and 2-forms* ([1]). – If $L_\nu^\mu$ is an element of $O(1, 3)$ then it, along with its inverse $\tilde{L}_\nu^\mu$, will act upon any frame $\mathbf{e}_\mu$ on Minkowski space and its reciprocal coframe $\theta^\mu$ :

$$\overline{\mathbf{e}}_\mu = \mathbf{e}_\nu \tilde{L}_\mu^\nu, \qquad\qquad \overline{\theta}^\mu = L_\nu^\mu \, \theta^\nu \,. \qquad\qquad (3.1)$$

Those matrices and their inverses will then act upon the components of bivectors and 2-forms by way of:

$$\overline{\mathfrak{H}}^{\mu\nu} = L_\kappa^\mu L_\lambda^\nu \, \mathfrak{H}^{\kappa\lambda}, \qquad \overline{F}_{\mu\nu} = \tilde{L}_\mu^\kappa \, \tilde{L}_\nu^\lambda \, F_{\kappa\lambda} \,. \qquad (3.2)$$

One assumes that $L_\nu^\mu$ and its inverse belong to differentiable curves through the identity matrix, as above, with:

$$\omega_\nu^\mu = \frac{dL_\nu^\mu}{ds}\bigg|_{s=0} = -\frac{d\tilde{L}_\nu^\mu}{ds}\bigg|_{s=0} \,. \qquad\qquad (3.3)$$

When one differentiates both actions in (3.2), one will get:

$$\delta \mathfrak{H}^{\mu\nu} \equiv \frac{d\mathfrak{H}^{\mu\nu}}{ds}\bigg|_{s=0} = \omega_\kappa^\mu \, \delta_\lambda^\nu \, \mathfrak{H}^{\kappa\lambda} + \delta_\kappa^\mu \, \omega_\lambda^\nu \, \mathfrak{H}^{\kappa\lambda} = \omega_\kappa^\mu \, \mathfrak{H}^{\kappa\nu} - \omega_\lambda^\nu \, \mathfrak{H}^{\lambda\mu}, \qquad (3.4)$$

$$\delta F_{\mu\nu} \equiv \frac{dF_{\mu\nu}}{ds}\bigg|_{s=0} = -\omega_\mu^\kappa \, \delta_\nu^\lambda \, F_{\kappa\lambda} - \delta_\mu^\kappa \, \omega_\nu^\lambda \, F_{\kappa\lambda} = -\omega_\mu^\kappa \, F_{\kappa\nu} + \omega_\nu^\lambda \, F_{\lambda\mu} \,. \qquad (3.5)$$

If we lower and raise an index on $\mathfrak{H}$ and $F$, resp., then we will see that an application of the theorem above and its corollaries will say that we can also express these infinitesimal actions of the Lorentz group on bivectors and 2-forms in the form:

$$\delta \mathfrak{H}^{\mu\nu} = [\omega, \mathfrak{H}]^{\mu\nu}, \qquad \delta F_{\mu\nu} = -[\omega, F]_{\mu\nu} \,. \qquad (3.6)$$

Hence, one can just as well regard $F$ and $\mathfrak{H}$ as elements of $\mathfrak{so}(1, 3)$, although since they are dual objects, it would be better to regard $F$ as an element of $\mathfrak{so}(1, 3)^*$ and $\mathfrak{H}$ as an element of $\mathfrak{so}(1, 3)$. From (2.38), the difference essentially amounts to a sign.

*b. The time+space splitting.* – If one splits Minkowski space $\mathfrak{M}^4 = (\mathbb{R}^4, \eta)$ into a time line $[\mathbf{t}]$ and a spatial subspace $\Sigma$, so one has a direct sum decomposition:

$$\mathfrak{M}^4 = [\mathbf{t}] \oplus \Sigma, \qquad\qquad (3.7)$$

---

then one can also speak of an *adapted* frame for $\mathfrak{M}^4$, which amounts to a frame $\{\mathbf{e}_\mu, \mu = 0, \ldots, 3\}$ such that $\mathbf{e}_0$ generates $[\mathbf{t}]$ and $\{\mathbf{e}_i, i = 1, 2, 3\}$ spans $\Sigma$.

Dually, the decomposition (3.7) will imply a corresponding decomposition of $\mathfrak{M}^{4*}$:

$$\mathfrak{M}^4 = [t] \oplus \Sigma^*, \tag{3.8}$$

although this time, $[t]$ is composed of all linear functionals on $\mathfrak{M}^4$ that annihilate $\Sigma$, while the subspace $\Sigma^*$ is composed of all linear functionals that annihilate $\mathbf{t}$:

$$t(\mathbf{x}) = 0 \qquad \text{for all } \mathbf{x} \in \Sigma, \qquad\qquad \alpha(\mathbf{t}) = 0 \qquad \text{for all } \alpha \in \Sigma^*. \tag{3.9}$$

As a consequence, the reciprocal coframe $\{\theta^\mu, \mu = 0, \ldots, 3\}$ to $\mathbf{e}_\mu$ will be adapted to the decomposition (3.8). In particular, $\theta^0$ will generate $[t]$, and $\{\theta^i, i = 1, 2, 3\}$ will span $\Sigma^*$.

The decomposition (3.7) implies corresponding decompositions of the spaces of bivectors and 2-forms over $\mathfrak{M}^4$.

$$\Lambda_2 = [\mathbf{t}] \wedge \Sigma \oplus \Lambda_2\Sigma, \quad \Lambda^2 = [t] \wedge \Sigma^* \oplus \Lambda^2\Sigma. \tag{3.10}$$

Any 2-form that belongs to $[t] \wedge \Sigma^*$ will annihilate any bivector that belongs to $\Lambda_2\Sigma$, and any 2-form that belongs to $\Lambda^2\Sigma$ will annihilate any bivector that belongs to $[\mathbf{t}] \wedge \Sigma$.

As a result of (3.10), any bivector $\mathbf{b}$ and any 2-form $b$ can be expressed uniquely in the form:

$$\mathbf{b} = \mathbf{t} \wedge \mathbf{a} + \mathbf{c}, \qquad b = t \wedge a + c, \tag{3.11}$$

in which $\mathbf{a} \in \Sigma$, $\mathbf{c} \in \Lambda_2\Sigma$, $a \in \Sigma^*$, and $c \in \Lambda^2\Sigma$.

If $\{\mathbf{e}_\mu, \mu = 0, \ldots, 3\}$ is an adapted frame for $\mathfrak{M}^4$ relative to the decomposition (3.7) and $\{\theta^\mu, \mu = 0, \ldots, 3\}$ is its reciprocal coframe then $\{\mathbf{e}_0 \wedge \mathbf{e}_i, i = 1, 2, 3\}$ will be a frame for $[\mathbf{t}] \wedge \Sigma$, $\{\varepsilon_{ijk} \mathbf{e}_j \wedge \mathbf{e}_k, i, j, k = 1, 2, 3\}$ will be a frame for $\Lambda_2\Sigma$, $\{\theta^0 \wedge \theta^i, i = 1, 2, 3\}$ will be a coframe for $[t] \wedge \Sigma^*$, and $\{\varepsilon^{ijk} \theta^j \wedge \theta^k, i = 1, 2, 3\}$ will be a coframe for $\Lambda^2\Sigma$. Hence, if we use $\mathbf{e}_0$ for $\mathbf{t}$ and $\theta^0$ for $t$ then the decompositions in (3.11) can be expressed in the form:

$$\mathbf{b} = a^i \mathbf{e}_0 \wedge \mathbf{e}_i + \tfrac{1}{2} c^{ij} \mathbf{e}_i \wedge \mathbf{e}_j, \qquad b = a_i \theta^0 \wedge \theta^i + \tfrac{1}{2} c_{ij} \theta^i \wedge \theta^j. \tag{3.12}$$

In particular, the electromagnetic excitation bivector $\mathfrak{H}$ and the field strength 2-form $F$ can be expressed as:

$$\mathfrak{H} = \mathbf{e}_0 \wedge \mathbf{D} + \mathbf{H}, \qquad F = \theta^0 \wedge E - B, \tag{3.13}$$

with:

$$\mathbf{D} = D^i \mathbf{e}_i, \qquad \mathbf{H} = \tfrac{1}{2} H^{ij} \mathbf{e}_i \wedge \mathbf{e}_j, \qquad E = E_i \theta^i, \qquad B = \tfrac{1}{2} B_{ij} \theta^i \wedge \theta^j. \tag{3.14}$$

In order to relate the spatial bivector $\mathbf{H}$ to the usual vector that goes by that name and the spatial 2-form $B$ to the spatial vector $\mathbf{B}$, one must first define a volume element on $\mathfrak{M}^4$ and then split it into a temporal and spatial part. Since we already have an adapted frame and coframe, we define the 4-form:

$$V = \theta^0 \wedge \theta^1 \wedge \theta^2 \wedge \theta^3 = \frac{1}{4!} \varepsilon_{\kappa\lambda\mu\nu} \theta^\kappa \wedge \theta^\lambda \wedge \theta^\mu \wedge \theta^\nu, \qquad (3.15)$$

which can then be split into:

$$V = \theta^0 \wedge V_s, \qquad V_s = \theta^1 \wedge \theta^2 \wedge \theta^3 = \frac{1}{3!} \varepsilon_{ijk} \theta^i \wedge \theta^j \wedge \theta^k. \qquad (3.16)$$

The spatial 3-form $V_s$ then defines a spatial volume element.

One can then define the *Poincaré isomorphism* $\# : \Lambda_k \to \Lambda^{4-k}$, $k = 0, \ldots, 4$, which takes any $k$-vector $\mathbf{a}$ to the $4-k$-form:

$$\#\mathbf{a} = i_\mathbf{a} V. \qquad (3.17)$$

Similarly, the spatial volume element $V_s$ defines a Poincaré isomorphism $\#_s : \Lambda_k \Sigma \to \Lambda^{4-k} \Sigma$, $k = 0, 1, 2, 3$, which takes $\mathbf{a}$ to:

$$\#_s \mathbf{a} = i_\mathbf{a} V_s. \qquad (3.18)$$

Under that spatial isomorphism, vectors go to 2-forms and bivectors go to 1-forms. In particular, one can define the vector $\mathbf{B}$ and the 1-form $H$ that make:

$$B = \#_s \mathbf{B}, \qquad H = \#_s \mathbf{H}. \qquad (3.19)$$

That explains where the usual vector fields of Maxwell's equations come from, when one adds that typically if one is dealing with $\mathbb{R}^3$ and the Euclidian metric (whose components in an orthonormal frame will be $\delta_{ij}$) then one will have "accidental" isomorphisms between the four three-dimensional vector spaces that consist of $\mathbb{R}^3$, its dual, its space of bivectors, and its space of 2-forms. When one does not distinguish between contravariant and covariant indices, they can all be represented by vectors, in effect. Of course, one must distinguish between "polar" vectors, which belong to $\mathbb{R}^3$ or its dual, and "axial" vectors, which are really bivectors or 1-forms, and will involve the volume element in their transformations. One then sees from (3.13) that there is something more fundamental about regarding $B$ as a 2-form and $\mathbf{H}$ as a bivector field.

When one applies the time+space splitting to the basic isomorphisms that we have established of $\Lambda_2$ with $\mathfrak{so}(1, 3)$ and $\Lambda^2$ with $\mathfrak{so}(1, 3)^*$, one will see that the bivectors of the form $\mathbf{e}_0 \wedge \mathbf{D}$ correspond to infinitesimal boosts, while the ones of the form $\mathbf{H}$ will corresponding to infinitesimal rotations. Dually (as we shall see in the next subsection), the 2-forms of the form $\theta^0 \wedge E$ correspond to forces, while the ones of the form $B$ correspond to torques.

One can make this explicit by introducing the standard basis matrices $J_i$ and $K_i$ for $\mathfrak{so}(1, 3)$:

$$J_1 = \begin{bmatrix} 0 & 0 & 0 & 0 \\ 0 & 0 & 0 & 0 \\ 0 & 0 & 0 & 1 \\ 0 & 0 & -1 & 0 \end{bmatrix}, \quad J_2 = \begin{bmatrix} 0 & 0 & 0 & 0 \\ 0 & 0 & 0 & -1 \\ 0 & 0 & 0 & 0 \\ 0 & 1 & 0 & 0 \end{bmatrix}, \quad J_3 = \begin{bmatrix} 0 & 0 & 0 & 0 \\ 0 & 0 & 1 & 0 \\ 0 & -1 & 0 & 0 \\ 0 & 0 & 0 & 0 \end{bmatrix}, \tag{3.20}$$

$$K_1 = \begin{bmatrix} 0 & 1 & 0 & 0 \\ 1 & 0 & 0 & 0 \\ 0 & 0 & 0 & 0 \\ 0 & 0 & 0 & 0 \end{bmatrix}, \quad K_2 = \begin{bmatrix} 0 & 0 & 1 & 0 \\ 0 & 0 & 0 & 0 \\ 1 & 0 & 0 & 0 \\ 0 & 0 & 0 & 0 \end{bmatrix}, \quad K_3 = \begin{bmatrix} 0 & 0 & 0 & 1 \\ 0 & 0 & 0 & 0 \\ 0 & 0 & 0 & 0 \\ 1 & 0 & 0 & 0 \end{bmatrix}. \tag{3.21}$$

Under the association:

$$\mathbf{e}_0 \wedge \mathbf{e}_i \mapsto K_i, \qquad \mathbf{e}_i \wedge \mathbf{e}_j \mapsto \varepsilon_{ijk} J_k, \tag{3.22}$$

one gets a linear isomorphism $\Lambda_2 \cong \mathfrak{so}(1, 3)$ with the desired association of $\mathbf{D}$ with an infinitesimal boost and $\mathbf{H}$ with an infinitesimal rotation.

Note that if one includes a factor of 1/2 in the Cartan-Killing form:

$$<a, b> = \tfrac{1}{2}\mathrm{Tr}\, ab \tag{3.23}$$

and introduces the general notation for the basis elements $\{\varepsilon_a, a = 1, \ldots, 6\}$ with $\varepsilon_a = K_a$ for $a = 1, 2, 3$ and $\varepsilon_a = J_{a-3}$ for $a = 4, 5, 6$ then:

$$<\varepsilon_a, \varepsilon_b> = \eta_{ab} = \mathrm{diag}\ [+1, +1, +1, -1, -1, -1]. \tag{3.24}$$

Hence, the basis $\varepsilon_a$ for $\mathfrak{so}(1, 3)$ is orthogonal for a scalar product of signature type (3, 3) that is defined by the Cartan-Killing form. As we shall see in the next section, there is a natural scalar product on $\Lambda_2$ that makes the linear isomorphism $\Lambda_2 \cong \mathfrak{so}(1, 3)$ an isometry of orthogonal spaces.

The dual isomorphism $\Lambda^2 \cong \mathfrak{so}(1, 3)^*$ can be defined by using the basic matrices $J_i$, $K_i$, $i = 1, 2, 3$ as a reciprocal basis for $\mathfrak{so}(1, 3)^*$, although one must prefix the $J_i$'s with a minus sign in order to account for the fact that $< J_i, J_j > = - \delta_{ij}$. However, if one wishes to define a dual basis that is orthonormal for the Cartan-Killing form, as it gets defined on $\mathfrak{so}(1, 3)^*$, then one can simply use the same basis matrices $J_i$, $K_i$ with no change of sign.

*c. Work and energy.* – In mechanics, $\mathfrak{so}(1, 3)$ can represent infinitesimal "displacements" of points in Minkowski space, while the elements of its dual $\mathfrak{so}(1, 3)^*$ can represent "generalized forces." Hence, the bilinear pairing of an element $\delta\tilde{\mathbf{x}} \in \mathfrak{so}(1, 3)$ and an $F \in \mathfrak{so}(1, 3)^*$ will give a scalar:

$$\delta W = F(\delta \mathbf{x}) = F_\mu \; \delta x^\mu \tag{3.25}$$

that represents the virtual work that that is done by $F$ during the virtual displacement $\delta \mathbf{x}$.

If the electric field strength $E$ is to represent a force per unit charge (at least, in terms of units), and $\mathbf{D}$ is to represent a spatial density of electric dipole moments (which have the units of charge times distance, in their own right) then one can see that the scalar $E(\mathbf{D})$ should represent an energy density. This is probably why $\mathbf{D}$ was usually referred to as the "electric displacement," although it is probably empirically more consistent to think of it in terms of the response of a dielectric medium to the imposition of $E$; i.e., the electric excitation of the medium by the formation of electric dipoles.

Similarly, if $H$ has the units of torque per unit dipole and $\mathbf{B}$ has the units of a spatial density of magnetic dipoles then the scalar $H(\mathbf{B})$ will have the units of torque density, or really energy density, since the torque is acting through a unitless angle of rotation.

All of this leads credence to the idea that we can think of the 2-form $F$ as a generalized relativistic force that is associated with an element of $\mathfrak{so}(1, 3)^*$, while $\mathfrak{H}$ is a generalized displacement that is associated with an element of $\mathfrak{so}(1, 3)$, in such a way that $F(\mathfrak{H}) = F_{\mu\nu} \; \mathfrak{H}^{\mu\nu}$ will become an energy density. In fact, it is typically used as the Lagrangian density for the action functional that will give Maxwell's equations as the Euler-Lagrange equations.

If one decomposes $F$ and $\mathfrak{H}$ according to a time+space decomposition, as in (3.13), then one will get a corresponding decomposition of $F(\mathfrak{H})$:

$$F(\mathfrak{H}) = (\theta^0 \wedge E)(\mathbf{e}_0 \wedge \mathbf{D}) + (\theta^0 \wedge E)(\mathbf{H}) - B(\mathbf{e}_0 \wedge \mathbf{D}) \; - B(\mathbf{H}), \tag{3.26}$$

which will give:

$$F(\mathfrak{H}) = E(\mathbf{D}) - B(\mathbf{H}) = E_i \, D^i - B_i \, H^i, \tag{3.27}$$

when $\theta^0 \wedge E$ annihilates the space that $\mathbf{H}$ belongs to and $B$ annihilates the space that $\mathbf{e}_0 \wedge \mathbf{D}$ belongs to.

As we shall see shortly, (3.27) is not the most general expression that we can encounter in the electrodynamics of continuous media. In particular, $\mathbf{D}$ and $\mathbf{H}$ might each depend upon both $E$ and $B$, which would have the effect of making the middle two terms in (3.26) non-vanishing.

*d. Electromagnetic constitutive laws.* – The empirical side of electrodynamics generally gets introduced into Maxwell's equations by way of an electromagnetic constitutive law, which associates the excitation bivector $\mathfrak{H}$ with the field strength 2-form $F$ in a manner that is based upon the nature of the medium in which the fields are presumed to exist. The most common such laws (at least, for linear media) take the form of linear isomorphisms of the fibers of $\Lambda^2 M$ with the corresponding fibers of $\Lambda_2 M$ at each point of space-time $M$.

In order to get from the most general case of an electromagnetic constitutive law to a linear isomorphism, one must first note that 2-forms and bivector fields on $M$ actually belong to infinite-dimensional linear spaces of functions, so the most general association would take functions to functions. There are three basic types of operators on functions: algebraic, differential, and integral. Differential operators are not commonly used as electromagnetic constitutive laws, so we shall rule them out directly. Integral constitutive laws are associated with "dispersive" media, so if one wishes to consider only algebraic operators then one must be dealing with only non-dispersive media.

Among the algebraic operators, which are the ones that can be associated with maps of fibers of $\Lambda^2 M$ with fibers of $\Lambda_2 M$, the two basic types are linear and nonlinear. Certainly, it would be naïve to suggest that the nonlinear constitutive laws have no use in physics, since they are, in effect, the ultimate form of all constitutive laws empirically. However, for small enough field strengths, most electromagnetic media will behave linearly, so that is why one usually deals with the linear case first and then considers how things might change when the field strengths increase beyond the bounds in which the response of the medium is approximately linear.

Hence, when one is looking at a linear, non-dispersive electromagnetic medium, one can define a *(linear, non-dispersive) electromagnetic constitutive law* by an invertible map $C : \Lambda^2 M \to \Lambda_2 M$ such that:

1. $C$ takes the fiber of $\Lambda^2 M$ over $x \in M$ to the fiber of $\Lambda_2 M$ over that same point for all points of $M$. (Hence, it is a *vertical* map.)

2. $C_x : \Lambda_x^2 \to \Lambda_{2,x}$ is a linear isomorphism for each $x$.

One can further analyze the possible manifestations of linear, non-dispersive constitutive laws according to the way that the medium in question responds to the imposition of electric and magnetic fields, such as whether it is a conductor or insulator, which might even depend upon the type of fields, and whether electric or magnetic dipoles form. Often those two conditions are mutually exclusive: That is, magnetic materials are often conductors, while dielectrics are often insulators. Most optical materials tend to be non-magnetic insulators, such as the various glasses and quartz.

One then sees that what a linear, non-dispersive, electromagnetic constitutive law $C$ (or rather, its inverse $C^{-1}$) defines is a correlation between each vector space $\Lambda_{2,x}$ and its dual vector space $\Lambda_x^2$. If one introduces the notations ($i, j, k = 1, 2, 3$):

$$\mathbf{b}_i = \mathbf{e}_0 \wedge \mathbf{e}_i, \qquad \mathbf{b}^{i+3} = \tfrac{1}{2} \varepsilon^{ijk} \mathbf{e}_j \wedge \mathbf{e}_k, \qquad b^i = \theta^0 \wedge \theta^i, \qquad b_{i+3} = \tfrac{1}{2} \varepsilon_{ijk} \theta^j \wedge \theta^k \qquad (3.28)$$

then one can represent $\mathfrak{H}$ and $F$ in the forms:

$$\mathfrak{H} = D^i \mathbf{b}_i + H_i \mathbf{b}^{i+3}, \qquad F = E_i b^i - B^i \mathbf{b}_{i+3}, \qquad (3.29)$$

which also makes:

$$H^{ij} = \varepsilon^{ijk} H_k, \qquad\qquad B_{ij} = \varepsilon_{ijk} B^k. \qquad (3.30)$$

Of course, one must pause to note that, just as the fields $\mathfrak{H}$ and $F$ seem to be mixing field strengths with excitations, they also seem to be mixing contravariant objects with covariant ones.

The components of the linear isomorphism $C_x$ relative to the bases above on $\Lambda_{2, x}$ and $\Lambda_x^2$, with either four or two components, are defined by:

$$C_x (\theta^\kappa \wedge \theta^\lambda) = \tfrac{1}{2} C^{\kappa\lambda\mu\nu}(x) \, \mathbf{e}_\mu \wedge \mathbf{e}_\nu , \qquad C_x (b^a) = C^{ab}(x) \, \mathbf{b}_b , \qquad (3.31)$$

resp. Clearly, the matrix $C^{ab}$ seems more concise, although sometimes one might also want to keep track of the antisymmetries in the index-pairs $\kappa\lambda$ and $\mu\nu$.

One can put the matrix into $C^{ab}$ block form relative to a time+space decomposition of the fibers $\Lambda_{2, x}$ and $\Lambda_x^2$:

$$C^{ab} = \left[ \begin{array}{c|c} C^{ij} & C^i_{j+3} \\ \hline C^{i+3}_j & C_{i+3, j+3} \end{array} \right], \qquad (3.32)$$

which allows one to write the constitutive law as a system of linear equations:

$$\left. \begin{array}{l} D^i = C^{ij} E_j + C^i_{j+3} B^j , \\ H_i = C^j_{i+3} E_j + C_{i+3, j+3} B^j . \end{array} \right\} \qquad (3.33)$$

So far, there is no particular symmetry to the indices $ab$. However, one can polarize $C^{ab}$ accordingly:

$$C^{ab} = C^{ab}_+ + C^{ab}_- , \qquad C^{ab}_+ \equiv \tfrac{1}{2} (C^{ab} + C^{ba}), \qquad C^{ab}_- \equiv \tfrac{1}{2} (C^{ab} - C^{ba}). \qquad (3.34)$$

A further reduction is based upon the fact that the volume element already defines one linear isomorphism $\#^{-1} : \Lambda_x^2 \to \Lambda_{2, x}$ whose component matrix with respect to the chosen bases above is:

$$\#^{ab} = \left[ \begin{array}{c|c} 0 & \delta^i_j \\ \hline \delta^i_j & 0 \end{array} \right]. \qquad (3.35)$$

Hence, since this matrix is symmetric in $ab$, it will be included in $C^{ab}_+$, multiplied by a scalar factor $\alpha$. That means that we ultimately have a decomposition of $C^{ab}$ into:

$$C^{ab} = C^{ab}_0 + C^{ab}_- + \alpha \, \#^{ab}. \qquad C^{ab}_0 \equiv C^{ab}_+ - \alpha \, \#^{ab}. \qquad (3.36)$$

In the terminology of Hehl and Obukhov [5], the components $C^{ab}_0$, $C^{ab}_-$, $\alpha \#^{ab}$ of $C^{ab}$ are referred to as the *fundamental*, *skewon*, and *axion* parts of $C^{ab}$, respectively. The matrices of the first two can be put into the form:

$$C_0^{ab} = \begin{bmatrix} \varepsilon^{ij} & | & a_j^i \\ \hline a_i^j & | & -\tilde{\mu}_{ij} \end{bmatrix}, \qquad C_-^{ab} = \begin{bmatrix} b^{ij} & | & d_j^i \\ \hline -d_i^j & | & c_{ij} \end{bmatrix}, \qquad (3.37)$$

in which $\varepsilon_{ij}$ and $\tilde{\mu}_{ij}$ are symmetric, $b^{ij}$ and $c_{ij}$ are antisymmetric, and the matrices $a_j^i$ and $d_j^i$ have no particular symmetry. The matrices $\varepsilon_{ij}$ and $\tilde{\mu}_{ij}$ are referred to as the *dielectric strength tensor* and the inverse of the *magnetic permeability tensor* $\mu^{ij}$.

The reason for the minus sign before the $\tilde{\mu}_{ij}$ is to make the signature type of $C_0^{ab}$ consistent with that of $\#^{ab}$, which is (3, 3), although a frame that diagonalizes $\#^{ab}$ will not generally diagonalize $C_0^{ab}$. Indeed, when the medium is isotropic, in addition to non-dispersive and linear, $C_0^{ab}$ will take the form:

$$C_0^{ab} = \begin{bmatrix} \varepsilon\,\delta^{ij} & | & 0 \\ \hline 0 & | & -\dfrac{1}{\mu}\delta_{ij} \end{bmatrix}, \qquad (3.38)$$

and the equations in (3.33) will take the form:

$$\mathbf{D} = \varepsilon\,\mathbf{E}, \qquad \mathbf{B} = \mu\,\mathbf{H}, \qquad (3.39)$$

which is how they are usually presented in physics literature [**6, 7**].

The classical electromagnetic vacuum is then defined by making the medium homogeneous, in addition to everything else, so $\varepsilon_0$ and $\mu_0$ will then be assumed to be constants. Of course, that condition of constancy is frame-dependent, and the principle of Lorentz invariance only requires that the *product* $\varepsilon_0\,\mu_0$ must be independent of a choice of Lorentz frame, or rather:

$$c_0 = \frac{1}{\sqrt{\varepsilon_0\,\mu_0}}. \qquad (3.40)$$

It is indeed intriguing that introducing only a volume element on $\mathbb{R}^4$, but not a scalar product, will allow one to define a scalar product on $\Lambda_2$ and $\Lambda^2$ that has the same signature type as the Cartan-Killing form for $\mathfrak{so}(1, 3)$, which does require one to define a scalar product on $\mathbb{R}^4$. Of course, the scalar product that is defined by $C_0^{ab}$ has a more empirical nature in the eyes of physics, although it presumably has the same signature type as $\#^{ab}$. Hence:

**Theorem:**

*When one has defined a scalar product of the signature type* (3, 3) *on* $\Lambda_2$, *the linear isomorphism that takes bivectors in* $\Lambda_2$ *to infinitesimal Lorentz transformations in* $\mathfrak{so}(1,$

3) *will also be an isometry of orthogonal spaces relative to the Cartan-Killing form on* $\mathfrak{so}(1, 3)$.

Because the electromagnetic constitutive law of a medium in which electromagnetic fields exist is so fundamental to pre-metric electromagnetism, it would take us too far afield to give a more thorough discussion of the possibilities here, so we shall simply refer to the author's book on pre-metric electromagnetism for more details [**8**]. However, we shall mention that the linear isometry of $\Lambda_2$ with $\mathfrak{so}(1,3)$ also yields a linear isometry of $\Lambda^2$ with $\mathfrak{so}(1,3)^*$, when one gives $\Lambda^2$ a scalar product by way of $C_{0,ab}$, which is the inverse of $C_0^{ab}$, or $\#_{ab}$, which is the inverse of $\#^{ab}$.

As a result of the linear isomorphism $\iota : \Lambda_2 \to \mathfrak{so}(1,3)$, one can associate the electromagnetic constitutive law that takes $\Lambda^2$ to $\Lambda_2$ with a more mechanical constitutive law that associates $\mathfrak{so}(1,3)$ with $\mathfrak{so}(1,3)^*$. Actually, the inverse of the mechanical constitutive law $\mathfrak{C} : \mathfrak{so}(1,3) \to \mathfrak{so}(1,3)^*$ is easier to define directly:

$$\mathfrak{C}^{-1} = \iota \cdot C \cdot \iota^{\mathrm{T}}. \tag{3.41}$$

**4. Relationship to line geometry.** – In the previous installments of this series of articles [**1**], we have already discussed some of the aspects of how electromagnetic fields relate to line geometry, which were all based upon the fact that decomposable bivectors and 2-forms on $\mathbb{R}^4$ can represent lines in $\mathbb{R}P^3$. Now that we have shown how bivectors and 2-forms can represent infinitesimal Lorentz transformations, we can put the two ideas together and discuss how lines in $\mathbb{R}P^3$ can represent elements of $\mathfrak{so}(1, 3)$.

Actually, that topic is closely related to one of the oldest applications of line geometry, which is how lines in $\mathbb{R}P^3$ can represent rigid motions of Euclidian $\mathbb{R}^3$. That study went back to the ground-breaking treatise of Julius Plücker in 1868 [**9**], which was developed further by his student Felix Klein [**10**] and Eduard Study [**11**], and it was Plücker who introduced the notion of a "Dyname." To the French, the corresponding word was "torseur," and to the Englishman Sir Robert Ball [**12**], the kinematical object was a "screw" and the dual dynamical object was a "wrench." These ideas are still being applied by modern mechanical engineers, especially in the study of robot manipulators [**13**].

Since all of that is clearly rooted in Newtonian mechanics, it is useful to verify rigorously that the Lie algebra $\mathfrak{iso}(3)$ of infinitesimal rigid motions in three-dimensional Euclidian space is the Newtonian limit ($c \to \infty$) of the Lie algebra $\mathfrak{so}(1,3)$. Hence, we shall do that first, and then discuss how Lorentz transformations, as well as electromagnetic fields, can amount to screws and wrenches, at least, in the Newtonian limit, which should apply to the rest space of any measurer/observer.

We shall conclude by discussing the relationship between the Cartan-Killing form on $\mathfrak{so}(1, 3)$ and the Klein quadric on $\Lambda_2$.

*a.  Rigid motions as Newtonian limits of Lorentz transformations.* – In order to show how Lorentz transformations of $\mathfrak{M}^4$ will become rigid motions of $E^3$, it is entirely sufficient to show that not only are translations the Newtonian limits of boosts, but that situation also relates to the structure constants of the two algebras.  Here, it is better to express a boost in terms of $v$ and $c$ explicitly, instead of $\cosh \alpha$ and $\sinh \alpha$.

An elementary boost along the $x$-direction in $\mathbb{R}^2 = (t, x)$ to a frame with a relative velocity of $v$ with respect to the first one takes the form:

$$\overline{t} = \gamma \left( t + \frac{v}{c^2} x \right), \qquad \overline{x} = \gamma (vt + x), \qquad \gamma = \left( 1 - \frac{v^2}{c^2} \right)^{-1/2}. \tag{4.1}$$

We can then put this into matrix form:

$$\begin{bmatrix} \overline{t} \\ \overline{x} \end{bmatrix} = B(v) \begin{bmatrix} t \\ x \end{bmatrix}, \qquad B(v) \equiv \gamma \begin{bmatrix} 1 & v/c^2 \\ v & 1 \end{bmatrix}. \tag{4.2}$$

One can already see that the Newtonian limit of an elementary boost is a translation at the level of finite transformations:

$$\lim_{c \to \infty} B(v) \equiv \begin{bmatrix} 1 & 0 \\ v & 1 \end{bmatrix}. \tag{4.3}$$

It is important to see that, physically, this translation acts upon velocities, not points of space-time.

In order to get the infinitesimal generator of the boost $B(v)$, we make $v = v(s)$ a differentiable curve with $v(0) = 0$, which will then make $B(v(0)) = I$, and differentiate $B$ at $s = 0$:

$$b(a) = \frac{dB}{ds} \bigg|_{s=0} = \left\{ \dot{\gamma} \begin{bmatrix} 1 & v/c^2 \\ v & 1 \end{bmatrix} + \gamma \begin{bmatrix} 0 & \dot{v}/c^2 \\ \dot{v} & 0 \end{bmatrix} \right\}_{s=0} = a \begin{bmatrix} 0 & 1/c^2 \\ 1 & 0 \end{bmatrix}, \tag{4.4}$$

in which, we have, of course defined $a$ to be $\dot{v}(0)$.

It is already clear that since:

$$\lim_{c \to \infty} \begin{bmatrix} 0 & 1/c^2 \\ 1 & 0 \end{bmatrix} = \begin{bmatrix} 0 & 0 \\ 1 & 0 \end{bmatrix}, \tag{4.5}$$

the Newtonian limit of the elementary boost $b(1)$ will be an elementary translation in $\mathbb{R}^2$.

Hence, we represent the three elementary boosts in the $x$, $y$, and $z$ directions by the matrices ([1]):

---

([1])  We have replaced $1/c^2$ with $\lambda$ for the sake of brevity.  The Newtonian limit will then be the limit as $\lambda$ goes to zero.

$$b_1 = \begin{bmatrix} 0 & \lambda & 0 & 0 \\ 1 & & & \\ 0 & & 0 & \\ 0 & & & \end{bmatrix}, \quad b_2 = \begin{bmatrix} 0 & 0 & \lambda & 0 \\ 0 & & & \\ 1 & & 0 & \\ 0 & & & \end{bmatrix}, \quad b_3 = \begin{bmatrix} 0 & 0 & 0 & \lambda \\ 0 & & & \\ 0 & & 0 & \\ 1 & & & \end{bmatrix}, \qquad (4.6)$$

this time, instead of $K_i$, $i = 1, 2, 3$.

By direct calculation, one sees that the commutation relations for the set of basis elements $\{J_i, b_i, i = 1, 2, 3\}$ for the vector space $\mathfrak{so}(1, 3)$ are:

$$[J_i, J_j] = \varepsilon_{ijk} J_k \,, \qquad [b_i, J_j] = \varepsilon_{ijk} b_k \,, \qquad [b_i, b_j] = - \lambda \, \varepsilon_{ijk} b_k \,. \qquad (4.7)$$

These are clearly the commutation relations (i.e., structure constants) for $\mathfrak{so}(1, 3)$ when $\lambda = 1$, so the basis for the vector space $\mathfrak{so}(1, 3)$ also generates its Lie algebra when one takes commutator brackets of matrices.

It is also clear that in the Newtonian limit, only the last set of relations will change. In particular, the infinitesimal translations $a_i = b_i$ ($\lambda = 0$) that the $b_i$ go to when $\lambda = 0$ will commute with each other in the appropriate manner, and in fact, the limiting Lie algebra as $\lambda$ goes to 0 will be:

$$[J_i, J_j] = \varepsilon_{ijk} J_k \,, \qquad [a_i, J_j] = \varepsilon_{ijk} a_k \,, \qquad [a_i, a_j] = 0, \qquad (4.8)$$

which is that of $\mathfrak{iso}(3)$.

*b. Electromagnetic fields as motors.* – A general terminology for infinitesimal rigid motions and their dual forces and moments that evolved in engineering mechanics was that of referring to the infinitesimal rigid motion or the dual object as a *motor* (*mo*ment + vec*tor*) [**14**, **15**]. More precisely, the term "motor" usually referred to the combination of a force and a torque, although one can distinguish between infinitesimal rigid motions and their dual objects by referring to *kinematical* and *dynamical* motors accordingly. Hence, from our discussion above, we can also think of bivectors as kinematical motors and 2-forms as dynamical ones.

A particular type of motor that gets a lot of attention in engineering mechanics is the "screw," which is essentially a canonical form for an infinitesimal rigid motion. The dual object is then a "wrench," which is then a canonical form for a dynamical motor. The two theorems upon which the theory of screws (dynames, torsors) is based are Chasles's theorem, which is kinematical in character, and Poinsot's theorem, which has a dynamical character. (Historically, the latter preceded the former, though.) We shall begin by simply stating them and refer the interested reader to the literature (which is vast) for more details (cf., e.g., [**16**]):

**Chasles's theorem:**

*When a rigid body moves freely in space, one can represent any rigid motion of that body as a translation along an axis and a rotation about that axis.*

**Poinsot's theorem:**

*When a finite set of discrete forces acts upon various points of a rigid body in space, that system of forces can be combined into a single force that acts along a line and a force-moment about that line.*

In both cases, if we understand "space" to actually mean $\mathbb{R}P^3$ then we are saying that there is always a "canonical form" for a rigid motion or a finite system of discrete forces that consists of a line in space, which is called the *central axis* of the motion or system of forces, and a plane that is perpendicular to that line. In the kinematical case, the line is the axis of both translation and rotation, while the plane is the plane of rotation; in the dynamical case, the line is the line of force, as well as the axis of the force-moment, and the plane is the plane of the force-moment ($M = r \wedge F$). Notice that insofar as planes are dual to lines in the eyes of projective geometry, the canonical forms that we have described both involve a line in space and a dual object to a line.

The canonical form of a rigid motion is what is most commonly referred to as a *screw* nowadays, although Ball's terminology appeared somewhat later than the foundational work of Plücker, Klein, and Study. The use of the term "screw" is consistent with the fact that an initial point in space that is then subjected to a simultaneous translation along an axis and a rotation about it will describe a helix. Dually, the canonical form is called (by Ball) a *wrench*, since a massive point that is subjected to a simultaneous force along a line of action and torque about it will also describe a helix as a response. However, the term "wrench" is rarely used now, as opposed to "force screw" [**15**].

Since the velocity vector **v** and the angular velocity vector $\boldsymbol{\varpi}$ (= the vector associated with the 1-form $\omega = \#_s \boldsymbol{\omega}$ by means of the metric) are parallel for a screw, one can express both in the form $v\,\mathbf{u}$ and $\varpi\,\mathbf{u}$ for some unit vector along the central axis. The ratio:

$$\rho = \pm \left| \frac{\varpi}{v} \right| \tag{4.9}$$

is referred to as the *parameter* of the screw; the sign is positive iff **v** and $\boldsymbol{\varpi}$ point in the same direction and negative otherwise. Hence, $\rho = 0$ for a pure translation and the parameter will become infinite for a pure rotation.

Typically, the first problem associated with screws is finding the central axis. Say one is given an infinitesimal rigid motion $\mathbf{v} + \boldsymbol{\omega}$ that consists of a velocity vector **v** at a point in space and an angular velocity $\boldsymbol{\omega}$. An axis of rotation will be associated with the real eigenvalue (= 0) of the linear transformation $\boldsymbol{\omega}$. However, one is still free to parallel-translate that axis throughout space. The central axis will have the property that **v** lies in the plane that is perpendicular to the position vector **r** from any point on the axis to the point of application of **v**; one finds the axis from that property. Since $\mathbf{v}_t = \boldsymbol{\omega}\mathbf{r}$ is perpendicular to the plane of **r** and the central axis, the vector $\mathbf{v}_n = \mathbf{v} - \mathbf{v}_t$ will be parallel to the central axis. Hence, if one is given the canonical form $\mathbf{v}_n + \boldsymbol{\omega}$ for the infinitesimal rigid motion then one can reconstruct **v** from $\mathbf{v}_n + \boldsymbol{\omega}\mathbf{r}$.

The way that all of this relates to electromagnetic fields becomes clearer when we appeal to the isomorphisms $\Lambda_2 \cong \mathfrak{so}(1, 3)$ and $\Lambda^2 \cong \mathfrak{so}(1, 3)^*$ that we described above, in conjunction with the Newtonian limit that turns Lorentz transformations into to rigid motions. Hence, one can think of a general electromagnetic excitation bivector field $\mathfrak{H}$ as an association of a kinematical motor to each point of space-time and a general electromagnetic field strength 2-form $F$ as associating a dynamical one. At each point of space-time, one can then associate $\mathbf{D}$ with an infinitesimal translation, $\mathbf{H}$ with an infinitesimal rotation, $E$ with a force, and $B$ with a torque, as was suggested above. However, only a special class of electromagnetic fields will associate the same motor to every point, namely, spatially-constant ones. Hence, it is probably better to think of the electromagnetic fields as being associated with "local motors" that act upon the fibers of the bundles $\Lambda_2(M)$ and $\Lambda^2(M)$ of bivectors and 2-forms on the space-time manifold $M$. More generally, one would essentially have to drop the constraint of rigidity of the moving body in order to get a formal analogy between electromagnetic and mechanical notions.

Similarly, one sees that only special configurations of electric and magnetic fields can be associated with screws, namely, ones for which $\mathbf{D}$ is parallel to $\mathbf{H}$ or $E$ is parallel to $B$. More generally, one would have to speak of a local screw that pertains to some particular point of space-time.

*c. Cartan-Killing form and the Klein quadric.* – We have already mentioned the fact that the Cartan-Killing form on $\mathfrak{so}(1, 3)$ has the same signature type as either the scalar product on $\Lambda_2$ that is defined by the volume element $V$, namely:

$$<\mathbf{A}, \mathbf{B}> = V (\mathbf{A} \wedge \mathbf{B}) = \#(\mathbf{A})(\mathbf{B}) = \#(\mathbf{B})(\mathbf{A}), \qquad (4.10)$$

or the fundamental part $C_0$ of any non-dispersive, linear constitutive law $C : \Lambda^2 \rightarrow \Lambda_2$, which takes the form:

$$C_0 (F, G) = F(C_0(G)) = G(C_0(F)). \qquad (4.11)$$

One interesting aspect of that situation is that one can define the scalar product (4.10) without introducing any physically-empirical data, but in order to get to the Lorentz group in the pre-metric theory of electromagnetism, one must look at the dispersion law for electromagnetic waves that follows from the electromagnetic constitutive law after making numerous reductions in generality, such as making it non-dispersive, linear, and isotropic. Hence, it seems that the appearance of things that pertain to the Lorentz group in electromagnetism can be associated with a much more elementary level of geometric considerations than the ones that follow from the constitutive law. In particular, the appearance of the Cartan-Killing form for $\mathfrak{so}(1, 3)$ requires only that one restrict oneself to a four-dimensional vector space and give it a volume element. Of course, one will still require the Minkowski scalar product if one is to define the linear isomorphism of $\Lambda_2$ with $\mathfrak{so}(1, 3)$, when it is regarded as a vector space.

If we use the scalar product that is defined by $V$ as an example then since the frame $\{\mathbf{b}_a, a = 1, \ldots, 6\}$ on $\Lambda_2$ that we have been using does not make the associated scalar product diagonal, but the basis $\{J_i, K_i, i = 1, 2, 3\}$ that we are using on $\mathfrak{so}(1, 3)$ does make the Cartan-Killing form diagonal, the first thing that we have to do is to find a basis for $\Lambda_2$ that diagonalizes the scalar product. That is quite straightforward, and we define:

$$Z_i = \frac{1}{\sqrt{2}} (\mathbf{b}_i + \mathbf{b}_{i+3}), \qquad \overline{Z}_i = \frac{1}{\sqrt{2}} (\mathbf{b}_i - \mathbf{b}_{i+3}), \tag{4.12}$$

for which we will have:

$$<Z_i, Z_j> = \delta_{ij}, \qquad <Z_i, \overline{Z}_j> = 0, \qquad <\overline{Z}_i, \overline{Z}_j> = -\delta_{ij}. \tag{4.13}$$

Hence, the frame $\{Z_i, \overline{Z}_j\}$ is orthonormal for the scalar product. The linear isomorphism $\Lambda_2 \to \mathfrak{so}(1, 3)$ that takes $Z_i$ to $K_i$ and $\overline{Z}_i$ to $J_i$ will then also define an isometry of the scalar products that preserves the component matrices.

As discussed in an earlier article in this sequence [**1**.I], the quadric hypersurface in $\Lambda_2$ (or really its projective space $P\Lambda_2$) that is defined by all bivectors $\mathbf{A}$ such that:

$$<\mathbf{A}, \mathbf{A}> = 0 \tag{4.14}$$

is called the "Klein quadric," and a bivector is decomposable (i.e., $\mathbf{A} = \mathbf{a} \wedge \mathbf{b}$) iff it belongs to that quadric.

We then look for the elements $\omega \in \mathfrak{so}(1, 3)$ that correspond to the points of the Klein quadric in $\Lambda_2$ under the isomorphism in question, so:

$$<\omega, \omega> = \tfrac{1}{2} \mathrm{Tr} \, \omega^2 = 0. \tag{4.15}$$

We shall call such elements *isotropic* for the scalar product that is defined by the Cartan-Killing form. Note that the condition (4.15) is quadratic and homogeneous, as one would expect from a quadratic form, as is the condition for the Klein quadric. Hence, the isotropic elements of $\mathfrak{so}(1, 3)$ will form a quadric hypersurface in $\mathfrak{so}(1, 3)$, which will then be five-dimensional; we shall then call it the *Cartan-Killing quadric.*

If we perform essentially the same transformation of the basis $\{J_i, K_i, i = 1, 2, 3\}$ that we did on $\{\mathbf{b}_a, a = 1, \ldots, 6\}$, namely:

$$z_i = \frac{1}{\sqrt{2}} (J_i + K_i), \qquad \overline{z}_i = \frac{1}{\sqrt{2}} (J_i - K_i), \tag{4.16}$$

then we will get:

$$<z_i, z_j> = <\overline{z}_i, \overline{z}_j> = 0, \qquad <z_i, \overline{z}_i> = \delta_{ij}. \tag{4.17}$$

Hence, this basis for $\mathfrak{so}(1, 3)$ behaves like the original basis $\{\mathbf{b}_a, a = 1, \ldots, 6\}$ does for $\Lambda_2$.

In particular, we see that we already know six distinct points on the Cartan-Killing quadric, namely, the six basis elements $\{z_i, \overline{z}_i, i = 1, 2, 3\}$. They are then related by just the algebraic relations that take the form of (4.17). When we look at the geometric nature of the elements of that basis, we see that they each consist of an infinitesimal boost along and axis and an infinitesimal rotation about that same axis. In other words, the isotropic elements of $\mathfrak{so}(1, 3)$ that are proportional to $z_i$ and $\overline{z}_i$ are essentially "relativistic screws," while the corresponding elements of the dual quadric on $\mathfrak{so}(1, 3)^*$ will be "relativistic force screws."

If one has, in general:

$$\Omega = \omega^i J_i + v^i K_i \tag{4.18}$$

then since:

$$\langle \Omega, \Omega \rangle = v^2 - \omega^2, \tag{4.19}$$

the most general element of the Cartan-Killing quadric will be characterized by having $v = \pm \omega$. However, that does not imply that $\omega^i = v^i$, or even that the two vectors are parallel, so there are other null elements than relativistic screws.

Conversely, since the general relativistic screw will take the form:

$$\Omega = \alpha(\rho J_i + K_i), \tag{4.20}$$

one will have:

$$\langle \Omega, \Omega \rangle = \alpha^2 (1 - \rho^2), \tag{4.21}$$

which will vanish only when $\rho = \pm 1$. That is, not all relativistic screws are null elements.

*d. Almost-complex structure on $\Lambda_2$.* – If we return to the basis $\{\mathbf{b}_a, a = 1, \ldots, 6\}$ for $\Lambda_2$ and define a linear isomorphism $* : \Lambda_2 \to \Lambda_2$ by way of:

$$*\mathbf{b}_i = \mathbf{b}_{i+3}, \qquad *\mathbf{b}_{i+3} = -\mathbf{b}_i \tag{4.22}$$

then we will see that:

$$*^2 = -I. \tag{4.23}$$

Hence, $*$ defines an almost-complex structure on $\Lambda_2$.

The general bivector can then be written in the form:

$$\mathbf{A} = \alpha^i \mathbf{b}_i + \beta^i *\mathbf{b}_i. \tag{4.24}$$

The almost complex structure then allows us to define its conjugate by:

$$\overline{\mathbf{A}} = \alpha^i \mathbf{b}_i - \beta^i *\mathbf{b}_i. \tag{4.25}$$

If one considers the basis that is defined by (4.12) then one will see that $\overline{Z}_i$ is, in fact, the conjugate of $Z_i$ with this definition.

The scalar product of $\mathbf{A}$ and $\mathbf{B} = \sigma^i \mathbf{b}_i + \tau^i *\mathbf{b}_i$ then becomes:

$$\langle \mathbf{A}, \mathbf{B} \rangle = \delta_{ij} (\alpha^i \sigma^j + \beta^i \tau^j), \tag{4.26}$$

so:

$$\langle \mathbf{A}, \mathbf{A} \rangle = 2 \, \delta_{ij} \, \alpha^i \beta^j = 2 \, \boldsymbol{\alpha} \cdot \boldsymbol{\beta}, \tag{4.27}$$

which is consistent with the real form.

However, since the almost complex structure * is linear and self-adjoint, i.e.:

$$\langle \mathbf{A}, *\mathbf{B} \rangle = \langle *\mathbf{A}, \mathbf{B} \rangle, \tag{4.28}$$

it also allows one to define another scalar product:

$$(\mathbf{A}, \mathbf{B}) = \langle \mathbf{A}, *\mathbf{B} \rangle = \delta_{ij} (\alpha^i \sigma^j - \beta^i \tau^j), \tag{4.29}$$

for which:

$$(\mathbf{A}, \mathbf{A}) = \delta_{ij} (\alpha^i \alpha^j - \beta^i \beta^j) = \boldsymbol{\alpha}^2 - \boldsymbol{\beta}^2. \tag{4.30}$$

This is the other field invariant that is commonly used in electromagnetism, along with (4.27).

Analogous considerations on $\Lambda^2$ will give analogous results. In particular, the almost-complex structure * on $\Lambda^2$ will define a corresponding almost-complex structure on $\Lambda_2$ by way of:

$$*\alpha (\mathbf{A}) = \alpha (*\mathbf{A}). \tag{4.31}$$

As it turns out, the Hodge * operator defines an almost-complex structure for 2-forms on a four-dimensional Lorentzian manifold. However, that operator is the starting point for all pre-metric electromagnetism, since it is the only place where the metric structure of space-time is actually used in Maxwell's equations. Indeed, one replaces it with the composition of a (non-dispersive, linear) electromagnetic constitutive law and the Poincaré isomorphism #. However, as the author pointed out in [**17**], not all constitutive laws will actually yield something that is proportional to an almost-complex structure for that composition. One finds that although the scalar product (4.26) is general in scope, the scalar product (4.29) is closely tied to the Lorentzian structure that comes from the classical electromagnetic vacuum, and by way of the introduction of *.

*e. Complex structure on $\Lambda_2$.* – When an almost-complex structure * has been defined on $\Lambda_2$, in order to define a complex structure on $\Lambda_2$, all that one needs to do is to define complex scalar multiplication:

$$(\alpha + i\beta) \, \mathbf{A} = \alpha \, \mathbf{A} + \beta \, *\mathbf{A}. \tag{4.32}$$

That definition allows one to regard $\{\mathbf{b}_i, i = 1, 2, 3\}$ as a complex basis for $\Lambda_2$, since other basis elements $\mathbf{b}_{i+3}$ will then become simply $i \, \mathbf{b}_i$. Hence, the complex dimension of $\Lambda_2$ is three.

The general element of $\Lambda_2$ will then take the form:

$$\mathbf{A} = (\alpha^i + i\beta^i) \, \mathbf{b}_i, \tag{4.33}$$

and a relativistic screw will take the form:

$$\mathbf{A} = \alpha\,(1 + i\rho)\,\mathbf{b} \tag{4.34}$$

for some bivector $\mathbf{b}$ that generates the central axis.

One can define the complex conjugate of a bivector (4.33) in the obvious way:

$$\overline{\mathbf{A}} = (\alpha^i - i\beta^i)\,\mathbf{b}_i, \tag{4.35}$$

and the basis (4.12) will take the form:

$$Z_i = \frac{1}{\sqrt{2}}\,(1 + i)\,\mathbf{b}_i, \qquad \overline{Z}_i = \frac{1}{\sqrt{2}}\,(1 - i)\,\mathbf{b}_i, \tag{4.36}$$

which is consistent with the definition of a complex conjugate, since $\mathbf{b}_i$ is a real bivector,

Notice that if we wish to introduce a scalar product on the complex form of $\Lambda_2$ that corresponds to $\mathbf{A} \wedge \mathbf{B}$ then we must first note that the exterior product is no longer defined on $\Lambda_2$ with its complex structure. Rather, if:

$$\mathbf{A} = (\alpha^i + i\beta^i)\,\mathbf{b}_i, \qquad \mathbf{B} = (\sigma^i + i\tau^i)\,\mathbf{b}_i \tag{4.37}$$

then we define their *complex* scalar product by making $\mathbf{b}_i$ orthonormal for the three-dimensional Euclidian scalar product, so:

$$<\mathbf{A},\,\mathbf{B}>_{\mathbb{C}} = \delta_{ij}\,(\alpha^i + i\beta^i)\,(\sigma^j + i\tau^j) = \delta_{ij}\,(\alpha^i\sigma^j - \beta^i\,\tau^j) + i\,\delta_{ij}\,(\alpha^i\,\tau^j + \beta^i\,\sigma^j). \tag{4.38}$$

That includes *both* of the scalar products that were defined above on $\Lambda_2$ with an almost-complex structure, since:

$$<\mathbf{A},\,\mathbf{B}>_{\mathbb{C}} = (\mathbf{A},\,\mathbf{B}) + i<\mathbf{A},\,\mathbf{B}>. \tag{4.39}$$

In particular:

$$<\mathbf{A},\,\mathbf{A}>_{\mathbb{C}} = (\boldsymbol{\alpha}^2 - \boldsymbol{\beta}^2) + i\,(2\,\boldsymbol{\alpha}\cdot\boldsymbol{\beta}), \tag{4.40}$$

which includes both of the commonly-used scalar products on bivectors. Hence, the null elements of this scalar product must satisfy both of the conditions:

$$0 = \boldsymbol{\alpha}^2 - \boldsymbol{\beta}^2, \quad 0 = \boldsymbol{\alpha}\cdot\boldsymbol{\beta}, \tag{4.41}$$

and when one interprets that in terms of electromagnetic fields, that will imply that one is dealing with perpendicular $\mathbf{D}$ and $\mathbf{H}$ fields that have the same magnitude, such as one would encounter with electromagnetic wave fields, but not exclusively.

Note that the quadric that $<.,.>_{\mathbb{C}}$ defines by its vanishing is a subspace of the Klein quadric, namely, its intersection with the quadric that is defined by the vanishing of $(.,.)$.

Once again, the almost-complex structure on $\Lambda^2$ that is induced by the one on $\Lambda_2$ will allow one to define a complex structure on $\Lambda^2$ in analogous way, with analogous dual results. In particular, one can define a complex Euclidian scalar product and a dual quadric.

One should be careful to distinguish the complex structure on $\Lambda_2$ that one gets from an almost-complex structure * from the *complexification* of $\Lambda_2$, which amounts to replacing the real components of bivectors with complex ones. In particular, the former complex vector space has a complex dimension of three, while the latter has a complex dimension of six. The author has pointed out in an article [**18**] that this implies a certain simplification of the complex formulation of relativity, which typically uses the complexification of $\Lambda_2$ and then restricts to a three-complex-dimensional subspace that is defined by the "self-dual" elements. Hence, it does not actually use all of the complexified bivectors and 2-forms.

*f. The isomorphism of $\Lambda_2$ with $\mathfrak{so}(3, \mathbb{C})$.* – When $\Lambda_2$ is given an almost-complex structure, and therefore a complex structure, there will then be a $\mathbb{C}$-linear isomorphism $\Lambda_2 \cong \mathfrak{so}(3, \mathbb{C})$ that associates the basis elements $\{\mathbf{b}_i$, $i = 1, 2, 3\}$ with the three elementary rotations in $E^3$:

$$I_1 = \begin{bmatrix} 0 & 0 & 0 \\ 0 & 0 & 1 \\ 0 & -1 & 0 \end{bmatrix}, \qquad I_2 = \begin{bmatrix} 0 & 0 & -1 \\ 0 & 0 & 0 \\ 1 & 0 & 0 \end{bmatrix}, \qquad I_3 = \begin{bmatrix} 0 & -1 & 0 \\ 1 & 0 & 0 \\ 0 & 0 & 0 \end{bmatrix}. \qquad (4.42)$$

However, the Lie algebra $\mathfrak{so}(3, \mathbb{C})$ is isomorphic to the Lie algebra $\mathfrak{so}(1, 3)$ by the association:

$$I_i \rightarrow J_i, \qquad i\, I_i \rightarrow K_i. \qquad (4.43)$$

Hence, one can simply regard an infinitesimal boost as an imaginary rotation.

The general element of $\mathfrak{so}(3, \mathbb{C})$ then takes the form:

$$\Omega = (\omega^i + i\, v^i)\, I_i. \qquad (4.44)$$

The Cartan-Killing form on $\mathfrak{so}(3, \mathbb{C})$ will be isometric to the Euclidian scalar product on $\mathbb{C}^3$, whose components will be $\delta_{ij}$ for a complex-orthonormal basis. One can also get this elementary fact directly from the fact that the Cartan-Killing form on $\mathfrak{so}(3, \mathbb{R})$ has that property in a real form by simply complexifying $\mathbb{R}^3$ to $\mathbb{C}^3$. The vector cross product on will then define the algebra of $\mathfrak{so}(3, \mathbb{C})$. Dually, the corresponding (transposed) scalar product on $\mathfrak{so}(3, \mathbb{C})^*$ will also make it isometric to complex three-dimensional Euclidian space.

A relativistic screw will take the form:

$$\Omega = v\,(\rho + i)\,\hat{\varpi}\,, \qquad\qquad (4.45)$$

for some real rotation $\hat{\varpi}$ with unit norm relative to the Cartan-Killing form, and some real numbers $v$ and $\rho$.

One can also associate null vectors and null covectors of the two complex Euclidian spaces $\mathfrak{so}(3, \mathbb{C})$ and $\mathfrak{so}(3, \mathbb{C})^*$ with the null vectors in $\Lambda_2$ and $\Lambda^2$ when they are regarded as three-dimensional complex Euclidian spaces. However, only the electromagnetic fields for which *both* real invariants $<.,.>$ and $(.,.)$ vanish (such as electromagnetic waves) in $\Lambda_2$ will go to null elements of $\mathfrak{so}(3, \mathbb{C})$, and only the analogous fields in $\Lambda^2$ will go to null elements of $\mathfrak{so}(3, \mathbb{C})^*$.

Note that the null elements of $\mathfrak{so}(3, \mathbb{C})$ and $\mathfrak{so}(3, \mathbb{C})^*$ include the elements of the form $J_i \pm iJ_i = (1 \pm i)\,J_i$, which amount to the combination of an infinitesimal rotation (torque, resp.) along a certain axis with an infinitesimal boost (force, resp.) along that same axis. Hence, it is proper to think of those elements as representing relativistic kinematical (dynamical, resp.) screws of a certain type, namely, the parameter of the screw would have to be $\pm 1$. Once again, the null elements are not all relativistic screws, nor are the relativistic screws all null, since from (4.45):

$$<\Omega, \Omega>_{\mathbb{C}} = v^2\,(\rho^2 - 1), \qquad\qquad (4.46)$$

which vanishes iff $\rho = \pm 1$.

The representation of electromagnetic fields by complex 3-vector fields goes back to the time of Riemann [**19**], and the technique was later developed by Ludwik Silberstein [**20**] and A. Conway [**21**]. It is closely related to the use of such fields to represent relativistic quantum wave functions by Majorana [**22**] and Oppenheimer [**23**]. The author has also discussed the role of complex structures in pre-metric electromagnetism [**24**]. Furthermore, the use of complex 3-vector fields is also established in the complex formulation of relativity [**25**], in which they are sometimes referred to as "3-spinors"; one might confer the author's own comments on that topic [**18**].

**5. Summary.** – In conclusion, we shall distill out the main points of the discussion above:

1. When bivectors and 2-forms are defined on four-dimensional Minkowski space, there are linear isomorphisms of $\Lambda_2$ with $\mathfrak{so}(1, 3)$ and $\Lambda^2$ with $\mathfrak{so}(1, 3)^*$, when the former is regarded as a vector space.

2. Under those isomorphisms, the electric excitation **D** gets associated with an infinitesimal boost, the magnetic excitation **H** goes to an infinitesimal rotation, the electric field strength $E$ becomes a force, and the magnetic field strength $H$ becomes a torque.

3. The bilinear pairing of 2-forms and bivectors that amounts to the evaluation of the 2-form on the bivector corresponds to the bilinear pairing of linear functionals in $\mathfrak{so}(1, 3)^*$ with elements of $\mathfrak{so}(1, 3)$.

4. The scalar products that one commonly defines on 2-forms and bivectors by way of a volume element or a linear, non-dispersive electromagnetic constitutive law have the same signature type as the Cartan-Killing form on $\mathfrak{so}(1, 3)$, and can make the linear isomorphisms in question into isometries.

5. One can then relate the electromagnetic constitutive law to a corresponding mechanical constitutive that associates elements of $\mathfrak{so}(1, 3)^*$ with elements of $\mathfrak{so}(1, 3)$.

6. The Lie algebra $\mathfrak{so}(1, 3)$ goes to the Lie algebra $\mathfrak{iso}(3)$ in the Newtonian limit as $c$ becomes infinite.

7. When bivectors and 2-forms are associated with lines in $\mathbb{R}P^3$, one can also associate elements of $\mathfrak{so}(1, 3)$ and $\mathfrak{so}(1, 3)^*$ with lines in a manner that reduces to the classical theory of motors in the Newtonian limit. However, only spatially-constant electromagnetic fields will correspond to a single infinitesimal rigid motion or dual object, so generally one must think of the motor as a local object that acts upon the fibers of the bundles of bivectors and 2-forms on space-time.

8. Relativistic screws are then elements of $\mathfrak{so}(1, 3)$ that consist of the sum of a boost along a central axis and a rotation about that same axis. Analogous considerations can be applied to $\mathfrak{so}(1, 3)^*$, which will then pertain to relativistic force screws.

9. The isotropic elements of $\mathfrak{so}(1, 3)$ with respect to the Cartan-Killing form – i.e., the points of the Cartan-Killing quadric – correspond to points of the Klein quadric in $\Lambda_2$.

10. The points of the Cartan-Killing quadric on $\mathfrak{so}(1, 3)$ include elements of the form $J_i \pm K_i$, which are essentially relativistic screws whose parameters are equal to $\pm 1$, but not all points of quadric are relativistic screws, nor are all relativistic screws found on that quadric. Analogous dual statements apply to $\mathfrak{so}(1, 3)^*$ and relativistic force screws.

11. When one introduces an almost-complex structure on the space of bivectors, one can also defined a complex structure on it, and the real linear isomorphisms of $\Lambda_2$ with $\mathfrak{so}(1, 3)^*$ and $\Lambda^2$ with $\mathfrak{so}(1, 3)^*$ will become $\mathbb{C}$-linear isomorphisms of $\Lambda_2$ with $\mathfrak{so}(3; \mathbb{C})$ and $\Lambda^2$ with $\mathfrak{so}(3; \mathbb{C})^*$, resp.

12. When $\Lambda_2$ has an almost-complex structure, one can define a complex Euclidian scalar product on it that combines both of the usual scalar products on $\Lambda_2$ that are defined in the Lorentzian formulation of electromagnetism. Therefore, the null vectors in $\mathfrak{so}(3; \mathbb{C})$ or $\mathfrak{so}(3; \mathbb{C})^*$ will correspond to the null vectors in $\Lambda_2$ or $\Lambda^2$, resp.

13. The Cartan-Killing form on $\mathfrak{so}(3; \mathbb{C})$ is the complex Euclidian metric, so the $\mathbb{C}$-linear isomorphisms $\Lambda_2 \cong \mathfrak{so}(3; \mathbb{C})$, $\Lambda^2 \cong \mathfrak{so}(3; \mathbb{C})^*$ are also isometries. Both of the scalar products on $\Lambda_2$ or $\Lambda^2$ must then vanish in order for the bivector or 2-form to be a complex null element. Electromagnetic fields that have that property include the fields of electromagnetic waves.

14. The vectors in $\mathfrak{so}(3; \mathbb{C})$ of the form $v\,(\rho + i)\,\hat{\varpi}$, where $\hat{\varpi}$ is some real rotation of unit norm, will represent special relativistic screws, while the corresponding null covectors $\mathfrak{so}(3; \mathbb{C})^*$ will represent special relativistic force screws.

————————

————————

([*])   References marked with an asterisk are available in English translation at the author's website: neo-classical-physics.info.

____________